\documentclass{article}

\usepackage{0_arxiv}

\usepackage[utf8]{inputenc} 
\usepackage[T1]{fontenc}    
\usepackage{hyperref}       
\usepackage{url}            
\usepackage{booktabs}       
\usepackage{wrapfig}        
\usepackage{amsmath}
\usepackage{amsfonts}       
\usepackage{mathrsfs}
\usepackage{mathtools}
\usepackage{dsfont}
\usepackage{nicefrac}       
\usepackage{microtype}      
\usepackage{todonotes}
\usepackage{accents}        
\usepackage{physics}
\usepackage{svg}
\usepackage{algpseudocode}
\usepackage{algorithm}
\newlength{\dhatheight}     

\renewcommand{\t}[1]{\text{#1}}

\DeclareMathOperator*{\argmax}{argmax}

\title{Investigation of inverse design of multilayer thin-films with conditional invertible Neural Networks}

\author{
  Alexander Luce \\
  University Erlangen-Nürnberg\\
  ams OSRAM\\
  Regensburg \\
\And
  Ali Mahdavi \\
  ams OSRAM\\
  Regensburg \\ 
\And
   Heribert Wankerl \\
   ams OSRAM\\
   Regensburg \\
\And
   Florian Marquardt \\
   University Erlangen-Nürnberg\\
   Max Planck Institute for the Science of Light\\
   Erlangen \\

}

\begin{document}
\maketitle

\begin{abstract}


The task of designing optical multilayer thin-films regarding a given target is currently solved using gradient-based optimization in conjunction with methods that can introduce additional thin-film layers. Recently, Deep Learning and Reinforcement Learning have been been introduced to the task of designing thin-films with great success, however a trained network is usually only able to become proficient for a single target and must be retrained if the optical targets are varied. In this work, we apply conditional Invertible Neural Networks (cINN) to inversely designing multilayer thin-films given an optical target. Since the cINN learns the energy landscape of all thin-film configurations within the training dataset, we show that cINNs can generate a stochastic ensemble of proposals for thin-film configurations that  that are reasonably close to the desired target depending only on random variables. By refining the proposed configurations further by a local optimization, we show that the generated thin-films reach the target with significantly greater precision than comparable state-of-the art approaches. Furthermore, we tested the generative capabilities on samples which are outside the training data distribution and found that the cINN was able to predict thin-films for out-of-distribution targets, too. The results suggest that in order to improve the generative design of thin-films, it is instructive to use established and new machine learning methods in conjunction in order to obtain the most favorable results.
\end{abstract}


\section{Introduction}
\label{sec:1_introduction}

In optics, being able to develop devices which manipulate light in a desired way is a key aspect for all applications within the field such as illumination \cite{Taki_2019} or integrated photonics \cite{nature:Pelucchi2022}. Recent developments in machine learning, deep learning and inverse design offer new possibilities to engineer such optical and photonic devices \cite{Hammond:19, Tahersima2019, Hedge2019, inverse_design, Ma2021-Deep_learning_for_the_design_of_photonic_structures, Peano_2021_PhysRevX.11.021052}. Nanophotonics in particular benefits from the recent advancements in optimization and design algorithms \cite{Hughes_2018_Inverse_Design_for_Nonlinear_Nanophotonic_Devices, Vuckovic_SPINS}. For example the development of meta optics or the design of scattering nano particles was greatly improved by employing gradient-based inverse design and deep learning \cite{Li2022-meta_optics, An:20-metasurface_deep_learning,Peurifoy-Nanophotonic_particle_inverse_design}. Multilayer thin-films are another instance of nanophotonic devices which are employed to fulfill a variety of different functionalities. Application examples are vertical-cavity surface-emitting lasers \cite{Iga_2018-VCSEL, Gebski:19}, anti-reflection coatings \cite{C1EE01297E-Anti_reflective_coatings} and wavelength demultiplexers \cite{Gerken:03-Multiplexing}. Recently, they were successfully employed to enhance the directionality of a white LED while maintaining the desired color temperature \cite{Wankerl2022-Directionality}. Designing multilayer thin-films \cite{Tikhonravov1993b, Becker2014, Tikhonravov:12} has been a task in the nanophotonics community for a long time and sophisticated techniques for the synthesis of thin-films, which exhibit desired optical characteristics have been developed in open-source or commercially available software \cite{optilayer, ThinFilmCenter, RPCoating, TFCalc, FilmWizard}. Methods such as the Fourier method \cite{Dobrowolski_Fourier_transforms, Larouche_needle_point} or the needle method \cite{Sullivan_needle_point, Larouche_needle_point, Tikhonravov_needle, Tikhonravov:12} compute the position inside the thin-film where the introduction of a new layer is most beneficial. Then the software will continue with a refinement process, often based on a gradient-based optimization such as the Levenberg-Marquardt algorithm \cite{levenberg-marquardt, Larouche_open_filters}, until it reaches a local minimum where it will then introduce another layer. Although the software will often converge to a satisfying solution with respect to the given target, the presented solutions often use excessive amounts of layers and the optimization is still limited by the selected parameters in the beginning of the optimization. The problem of converging to local optima was tackled in the past by the development of numerous global optimization techniques which have been introduced and tested in the field of thin-film optimization \cite{Chang1990, Paszkowicz2013, Yang2013, Guo2014, Martin1995}. Recently, the innovations of machine learning attracted much interest in the thin-film community and resulted in interesting new ways to create thin films \cite{Hedge2019, Roberts2018}. Particularly, deep reinforcement learning or Q-learning showed promising results in designing new and efficient multilayer thin-films while punishing complicated designs, which employ many layers \cite{Jiang2020, wankerl2020parameterized} and require targets that are difficult to achieve with conventional optimization.

In this work we employ so called conditional Invertible Neural Networks (cINNs) \cite{ardizzone2021conditional} to directly infer the loss landscape of all thin-film configurations with a fixed number of layers and material choice. The cINN learns to map the thin-film configuration to a latent space, conditional on the optical properties, ie. the reflectivity of a thin-film. During inference, due to the invertibility of the architecture, the cINN maps selected points from the latent space to their most likely thin-film configurations, conditional on a chosen target. This results in requiring only a single application of the cINN to obtain the most likely thin-film configuration given an optical target. Additionally, the log-likelihood training makes the occurrence of mode-collapse \cite{evaluation_mode_collapse} almost impossible. For thin-films, many different configurations lead to similar optical properties. For conventional optimization, this leads to the convergence of the optimization to possibly unfavorable local minima. A cINN circumvents this due to the properties of the latent space - by varying the points in the latent space, a perfectly trained cINN is able to predict any possible thin-film configuration that satisfies the desired optical properties. In this work, we investigated how good the generative capabilities of a cINN are for finding suitable thin-film configurations in a real-world application. We present an optimization algorithm, which is suitable to improve the thin-film predictions of the cINN. Then, we compared the optimization results of the presented algorithm to state-of-the-art software. Finally, we discuss the limitations of the approach and give a guideline when the application of a cINN is advantageous.



\section{Normalizing flows and conditional invertible neural networks}
\label{sec:3_theory_cINN}
Invertible neural networks are closely related to normalizing flows, which were first popularized by Dinh et. al. \cite{dinh2015nice}. A normalizing flow is an architecture that connects two probability distributions by a series of invertible transformations. The idea is to map a complex probability distribution to a known and simple distribution such as a Gaussian distribution. This can be used both for density estimation, but also for sampling since points can easily be sampled with a Gaussian distribution and mapped to the complex distribution via the normalizing flow. The architecture of a normalizing flow is constructed from the following. Assume two probability distributions, $\pi$ which is known and for which $z\sim \pi(z)$ holds and the complex, unknown distribution $p$. The mapping between both is given by the change-of-variables formula
\begin{align}
    p(x) = \pi(z) \left| \text{det}\left[\frac{\partial z}{\partial x} \right] \right|.
\end{align}
Consider a transformation $f$ which maps $f(x) = z$. Then the change-of-variables formula can be written as
\begin{align}
    p(x) = \pi(z) \left| \text{det}\left[\frac{\partial f(x)}{\partial x} \right] \right|.
\end{align}
The transformation $f$ can be given by a series of invertible transformations $f = f_K \circ f_{K-1} \ldots \circ f_0$ with $x = z_K = f(z_0) = (f_K\circ \ldots \circ f_0) (z_0)$. Then, the probability density at any intermediate point is given by $p_i(x_i) = z_i = f_i(z_{i-1})$. By rewriting the change-of-variables formula and taking the logarithm one obtains
\begin{align}
    \log{(p(x))} = \log{ \left(\pi(z_0) \prod_{i=1}^K \left| \text{det} \left[ \frac{\partial f_i(z_{i-1})}{\partial z_{i-1}} \right] \right|^{-1}\right)} = \log{(\pi(z_0))} - \sum_{i=1}^K \log{\left| \text{det} \left[ \frac{\partial f_i(z_{i-1})}{\partial z_{i-1}} \right] \right|}.
\end{align}
To be practical, a key component of any transformation of a normalizing flow is that the Jacobian determinant of the individual transformations must be easy to compute. A suitable invertible transformation, which is sufficiently expressive is the so called RNVP block \cite{dinh2017density}. The input $z$ is split into two separate vectors $u_1$ and $u_2$ and is processed with the help of two transformation functions, $s_2$ and $t_2$, while the other input $u_2$ is kept fixed
\begin{align}
    f_i(u_1 \otimes u_2) =  \{ u_1\odot\exp(s_2(u_2)) \oplus t_2(u_2)\otimes u_2\} = \{v_1 \otimes u_2\}.
\end{align}
$\odot, \oplus, \ominus$ and $\oslash$ denote element wise computation while $\otimes$ denotes a concatenation. The inverse of this transformation is given by
\begin{align}
    f_i^{-1}(v_1 \otimes u_2) &= \{ (v_1\ominus t_2(u_2))\oslash \exp(s_2(u_2)) \otimes u_2\}.
\end{align}
To invert the entire transformation, no inversion of the transformations $s_2$ and $t_2$ is required. Therefore, a neural network can be used as a transformation to make the normalizing flow expressive. The Jacobian of the transformation is given by an upper triangular matrix. Therefore, the Jacobian determinant is easy to compute since only the diagonal elements contribute.
\begin{align}
    \text{det} \left[\frac{\partial f_i}{\partial z}\right]  &= \text{det}
    \begin{bmatrix}
    \mathds{1}^{d} & 0\\
    \frac{\partial v_1}{\partial u_2} & \text{diag}(\exp(s_2(u_2) + t_2(u_2)))
    \end{bmatrix}.
\end{align}
Since only one part of the input is transformed by a neural network, the RNVP block is repeated and applied to the yet untransformed part of the input $u_2$ with two additional transformation functions $s_1$ and $t_1$. Other transformations, which can be advantageous depending on the specific application are the GLOW transformation \cite{kingma2018glow}, which utilizes invertible 1x1 convolutions or the masked autoregressive flow with a generalized implementation of the RNVP block \cite{masked_autoregeressive_flow_papamakarios}. 

Utilizing normalizing flows for the task of a generative neural network for the proposition of thin-films requires some modification of the normalizing flow. Ardizzone et. al. \cite{ardizzone2021conditional} propose a so called conditional invertible neural network (cINN) which extends the change-of-variables formula to conditional probability densities with condition c
\begin{align}
    p(x|c) = \pi(z) \left| \text{det}\left[\frac{\partial f(x;c)}{\partial x} \right] \right|.
\end{align}
By assuming a Gaussian probability distribution for $\pi$ and taking the logarithm, the conditional maximum likelihood loss function is derived by Ardizzone et. al. for training of a cINN as
\begin{align}
    \mathcal{L}_\text{cML} = \mathds{E}\left[ \frac{\norm{f(x;c)}_2^2}{2} - \log{\left| \text{det}\left[\frac{\partial f(x;c)}{\partial x} \right] \right|} \right].
\end{align}
The condition $c$ can be the result of another neural network $\varphi(c) = s$, which extracts features from the given condition $c$. The features are then passed to the RNVP transformations by appending the features to the input vectors $u_1$ and $u_2$. By jointly training the cINN via the conditional maximum likelihood loss, Ardizzone et. al. showed that the feature extraction network learns to extract useful information for the invertible transformation. The condition $c$ can be thought of as the target of the cINN. During inference, the cINN transforms gaussian samples to the learned distribution, conditional on the target $c$. Ardizzone et. al. also provide a Python implementation FrEIA\footnote{\url{https://github.com/VLL-HD/FrEIA\#papers}} for conditional and regular invertible neural network based on the Pytorch\footnote{\url{https://pytorch.org/}} Deep Learning library.

\section{Application of conditional invertible neural networks to generating multilayer thin-films}
\label{sec:4_application_to_thin-films}

\begin{figure}[t]
\centering
    \begin{minipage}{.54\textwidth}
    \centering
        \includegraphics[angle=0, trim = 0cm 0cm 0cm 0cm, clip, height=5cm,]{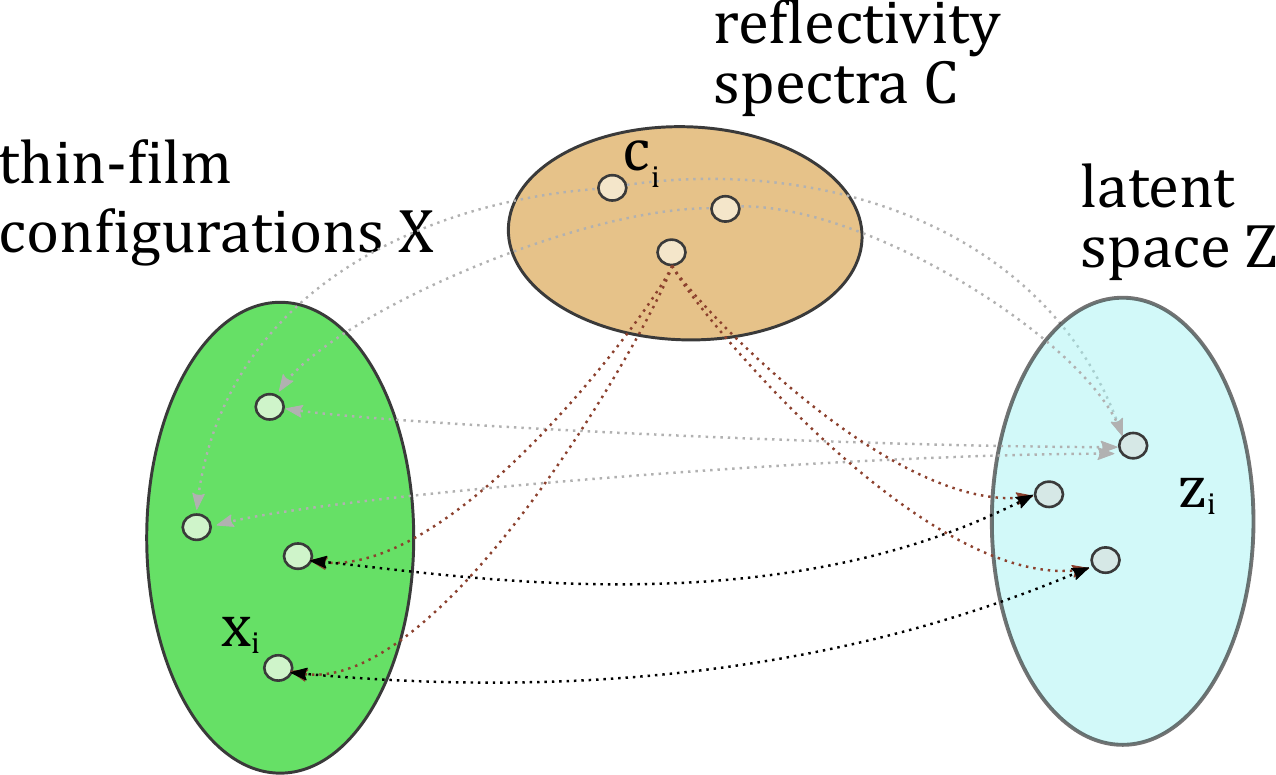}
        \caption{Visualization of the latent space. In the depiction, two thin-films from space $X$ have the same (or very similar) reflectivity, which is given via the condition in $C$. During training, they are mapped to different regions in the latent space $Z$. The latent space sample drawn during generation decides which solution is presented by the cINN. Other thin-films, which have a different reflectivity can be mapped to the same latent space point. They are distinguished by the condition i.e. their reflectivity. \label{fig:forward_with_latent_space}}
    \end{minipage}
    \hfill
    \begin{minipage}{0.4\textwidth}
    \centering
        \includegraphics[angle=0, trim = 0cm 0cm 0cm 0cm, clip, height=5cm]{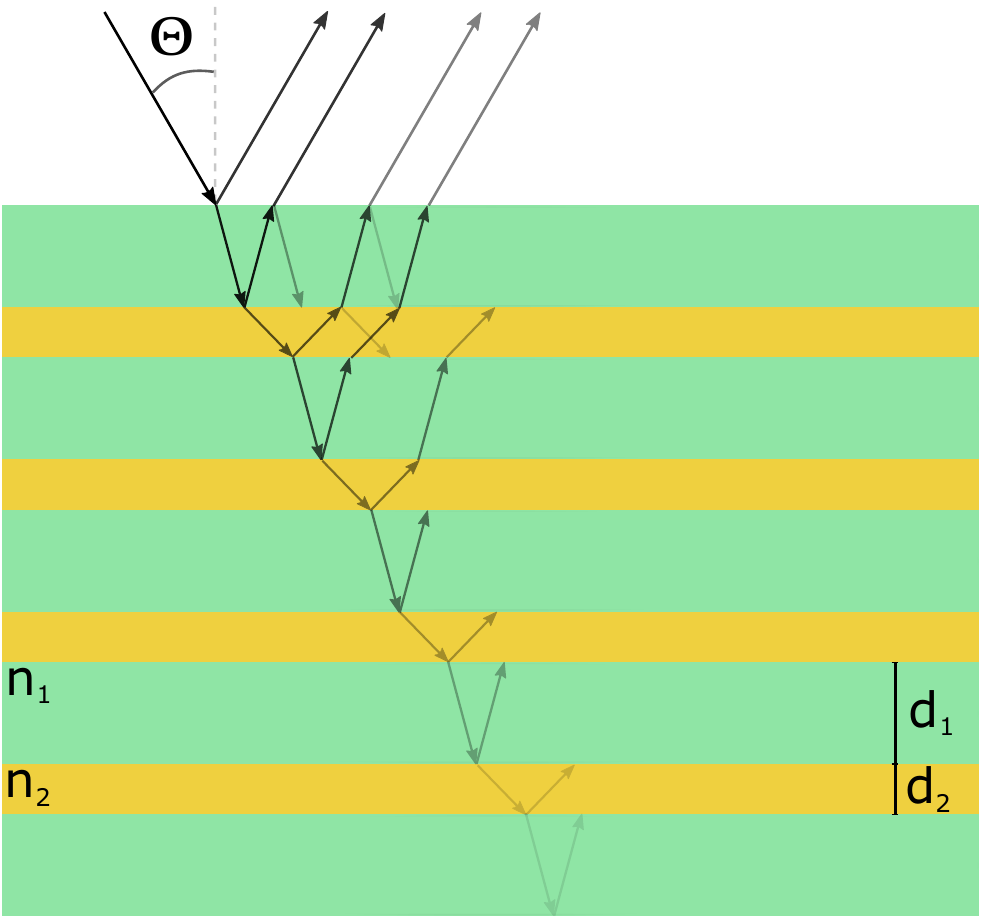}
        \caption{Schematic depiction of a multilayer thin-film. The two different materials have complex refractive indices $n_1$ and $n_2$. In this depiction, all layers with material $n_1$ have thickness $d_1$ and $n_2$ have thickness $d_2$, but the layer thicknesses can be chosen differently for any layer. The light is reflected and transmitted at every boundary and is also absorbed in the thin-film. \label{fig:thin-film_r_and_t}}
    \end{minipage}
\vspace{-.4cm}
\end{figure}

Multilayer thin-films are an optical component that consists of a sequence of planar layers with different materials stacked on top of each other with a varying layer thickness. Light, which irradiates the thin-film, can be transmitted and reflected at the layer interfaces and absorbed within the layer, as depicted in \autoref{fig:thin-film_r_and_t}. This interaction of reflection, transmission and absorption is different for different wavelengths of light and angles of incidence $\Theta$. A fast and convenient way to compute the optical response of a thin-film is to employ the transfer matrix method (TMM) \cite{byrnes2019multilayer}. In a previous work, we developed the Python package TMM-Fast\footnote{\url{https://github.com/MLResearchAtOSRAM/tmm_fast}} \cite{Luce:22} which allows to compute the optical response of a thin-film. The package also implements convenience functionality for thin-film-dataset generation, which is especially important for machine learning. By changing the layer thicknesses, the behavior of the thin-film can be modified, which are the parameters $\mathfrak{p}$ that are up to optimization. 

Finding a thin-film design, which fulfills the optical criteria while employing only a very limited number of layers is a challenging task since the loss landscape is highly non-convex and high-dimensional. The loss landscape is given by a mapping of the optical characteristics $\mathcal{M}$ of the thin-films by a loss function $\mathcal{L}(\mathcal{M}(\Theta, \lambda), \mathcal{M}_\text{target}(\Theta, \lambda)) = E$ with $E\in\mathds{R}$. The results of any local optimization, which is performed on the thin-film is highly dependent on the initial choice of the thin-film parameters as in any non-convex optimization. The idea of this paper is to interpret the loss landscape of the thin-films parameters as a probability distribution and train a conditional invertible neural network to find a mapping between the Gaussian latent space and the true loss landscape. The condition is then given by the optical characteristic of the thin-film $\mathcal{M}$. 

An important aspect of thin-films is also the possibility of ambiguity, which is a byproduct of the non-convex loss landscape. Two distinct thin-films $\mathfrak{p}_1 \neq \mathfrak{p}_2$ can have a very similar optical characteristic $\mathcal{M}_1 \approx \mathcal{M}_2$. Many generative models suffer under mode collapse \cite{evaluation_mode_collapse}, which slows the training of the neural network and often leads the neural network to ignore possible solutions or give wrong solutions altogether if ambiguity is present within the problem. The log-likelihood training of the cINN makes mode collapse virtually impossible to occur \cite{ardizzone2019analyzing} and turns ambiguous solutions distinguishable via the latent space, see \autoref{fig:forward_with_latent_space}. Therefore, it is possible to scan the latent space systematically for favorable design configurations.

\subsection{Multilayer thin-film dataset}\label{sec:datasets}
As noted before, the TMM-Fast package is used to create a dataset for training. In the presented case, we consider 9-layer thin-films and fix the material choice to an alternating sequence of niobium pentoxide (Nb2O5) and silicon dioxide (SiO2), while the light injection layer consists of gallium nitride (GaN) and the ambient outcoupling layer consists of air. For the dataset, one million thin-films are created for which the layer thickness of the 9 layers of Nb2O5 and SiO2 are uniformly sampled from a thickness range of $[10\t{nm}, 200\t{nm}]$. $10\,\%$ of the created thin-films are separated as test dataset. As optical response, we compute the reflectivity of the thin-film for normal incidence at 100 equidistant points in a range of $[400\t{nm}, 700\t{nm}]$. The computations were performed with our self-developed open-source transfer matrix solver \cite{Luce:22} and took about 40min on a Intel(R) Xeon(R) E5-2697v2 CPU. To test the generative capabilities, 600 so called target reflectivities are generated, which implement a selection of potentially interesting targets for which a priori no ground-truth thin-film with such an optical response is known. The dataset is normalized to mean $\mu=0$ and standard deviation $\sigma=1$ before it is processed by the neural networks. 

\subsection{Layout of the conditional invertible neural network}
As explained in \autoref{sec:3_theory_cINN}, a cINN consists of two neural networks, the invertible part and the feature extraction network, which processes the condition. The used feature extraction network is based on a ResNet architecture adapted to 1D convolutions. It features two residual shortcuts with 20-channel 1D convolutions, batch normalization, ReLU activation and a final dense layer with ReLU activation function. The feature extraction network receives the appropriate reflectivity spectrum (training or target), depending on whether the network is in training or inference mode. The output of the feature network is passed to the invertible neural network. The invertible part consists of eight "all\_in\_one\_blocks" from the FrEIA package. These blocks implement the RNVP transformation of the normalizing flow with additional functionality, such as active normalization \cite{kingma2018glow}. Here, a dense neural network with three layers, 512 neurons each and ReLU activations are used as secondary transformations for every RNVP block. Additionally, batch normalization is performed in the secondary transformations.

The training consists of 40 epochs of training, with an initial learning rate of 0.001 and a step decay of 0.08 of the learning rate after 20 and 25 epochs. We used Adam as optimizer with standard hyper-parameters. Training was performed on a GPU server with a Nvidia Tesla P100 graphics card and took less than one hour.

\subsection{Evaluation of the generative capabilities of the cINN}
\begin{figure}[t]
\centering
\includegraphics[angle=0, trim = 0cm 0.cm 0cm 0.cm, clip, width=0.85\textwidth]{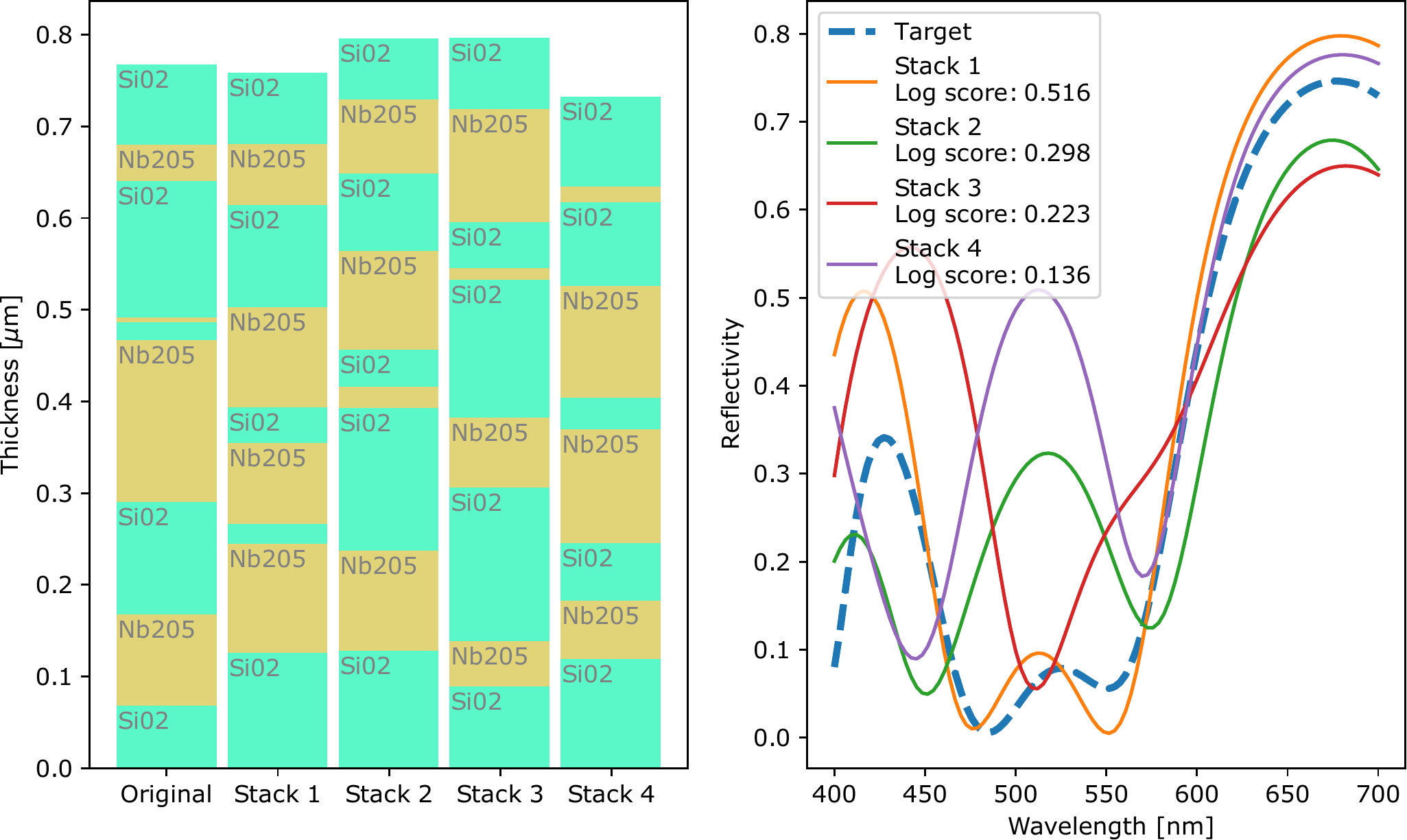}
\caption{Generation of thin-films via the cINN. A thin-film (Original) with its respective reflectivity is sampled from the test dataset. The reflectivity is used as the target for the cINN. Four different latent space points are sampled and evaluated by the cINN together with the extracted features of the target reflectivity by the feature extraction network. The four predicted thin-films (Stack 1-4) are depicted next to the original thin-film. They are then evaluated employing the tmm-fast package to validate if the prediction match the requested target. The best prediction is stack 1 with a log-score of $\mathcal{L} = 0.516$. Typically, a log-score of $>1$ is desired. \label{fig:test_generative_capabilities}}
\vspace{-.4cm}
\end{figure}

To test the generative capabilities of the trained cINN qualitatively, the reflectivity of a test dataset sample is taken and four random samples from the latent space are drawn. First, the reflectivity is processed by the feature extraction network. Then, the four drawn latent space samples and the extracted features are passed to the invertible network and processed by it. The generated thin-films (stack 1-4) are shown in \autoref{fig:test_generative_capabilities} together with the thin-film that generated the target during the dataset generation (labeled Original). To validate the performance of the generated thin-films, the TMM is used to compute the reflectivity. The performance metric displayed is the \textit{log-score} ($\mathcal{L}_\t{ls}$), which is the transformed mean squared error between the target and the validated reflectivity. The log-score is given by $\mathcal{L}_\t{ls} = -0.43 \log\left(\frac{1}{N}\sqrt{\sum_i (x_i - x_{\t{target,}\,i})^2}\right) - 0.52$. By performing a logarithmic transformation of the mean squared error the performance metric becomes more human-readable to quickly judge whether a thin-film is sufficiently approximating the target. The prefactors of the log-score are chosen arbitrarily in such a way that $\mathcal{L}_\t{ls}=1$ is a threshold below which a generated thin-film is not acceptable although log-scores $>$ 1 are still desired. The validated reflectivities of the propositions of the cINN capture the general trend of the target, they fail to match the target sufficiently and have a log-score of $<$ 1. Repeating the experiment with different targets yields similar results.

\subsection{Influence of the standard deviation on cINN predictions}
The latent space samples for the inverse application of the cINN are drawn from a Gaussian distribution. During training, the distribution of thin-films is mapped to a normal distribution (zero mean and standard deviation of one). Since the log-likelihood training preserves the probability density of the mapping between the thin-film distribution and the latent space, the log-probability of a latent space sample dictates the likelihood of the predicted thin-film to be the appropriate thin-film for the given condition. It is possible to think of the standard deviation as a "creativity parameter" for the cINN. Sampling with $\sigma\approx 0$ results in only the most likely thin-films while $\sigma>1$ results in explorative behavior of the cINN. Notably, the cINN will also predict nonphysical thin-films for latent space samples with low likelihood, e.g. layers with negative thickness. 

\subsection{Local optimization by Nelder-Mead downhill-simplex algorithm}
To improve upon the proposals from the cINN, a subsequent local optimization can be performed by using the suggested thin-film as initial starting values. The optimization is performed by using the downhill-simplex algorithm \cite{Nelder1965ASM} via the implementation of SciPy\footnote{https://docs.scipy.org/doc/scipy/reference/optimize.minimize-neldermead.html}. First, latent space values are sampled from a Gaussian distribution with zero mean and a standard deviation of one. The samples are passed to the invertible network together with the features extracted from the target. Then the thin-film with the best approximation is optimized via the downhill-simplex algorithm. The entire algorithm is also depicted as pseudo-code in algorithm \ref{alg:cinn_optimization}.
\begin{figure}[ht]
\vspace{-0.5cm}
      \begin{algorithm}[H]
        \begin{algorithmic}[1]
        \footnotesize
        \State Sample latent space values via z = $\mathcal{N}(0, \t{std})$
        \State Choose target t
        \State features = conditional\_network(t)
        \State proposals = invertible\_network(z, features)
        \State parameters = $\argmax(\mathcal{L}_\t{ls}(\t{proposals}))$
          \While{(iter < maxiter) and (error > tolerance)}
          \State parameters, error = downhill\_simplex\_step(parameters)
          \EndWhile \\
          \Return parameters, error
        \end{algorithmic}
         \caption{Pseudocode for the optimization of a thin-film with initialization by the cINN. The user needs to specify the \textit{loss function} $\mathcal{L}: \Vec{x}\mapsto l$, where $\Vec{x}$ denotes the optical properties of interest of the thin-film. $\mathcal{L}$ measures the performance of the thin-film such that the requirements are fulfilled if $l=0$.}
         \label{alg:cinn_optimization}
      \end{algorithm}
\end{figure}

\begin{figure}[ht]
\centering
\includegraphics[angle=0, trim = 0.cm .0cm 0.cm 0.cm, clip, width=0.85\textwidth]{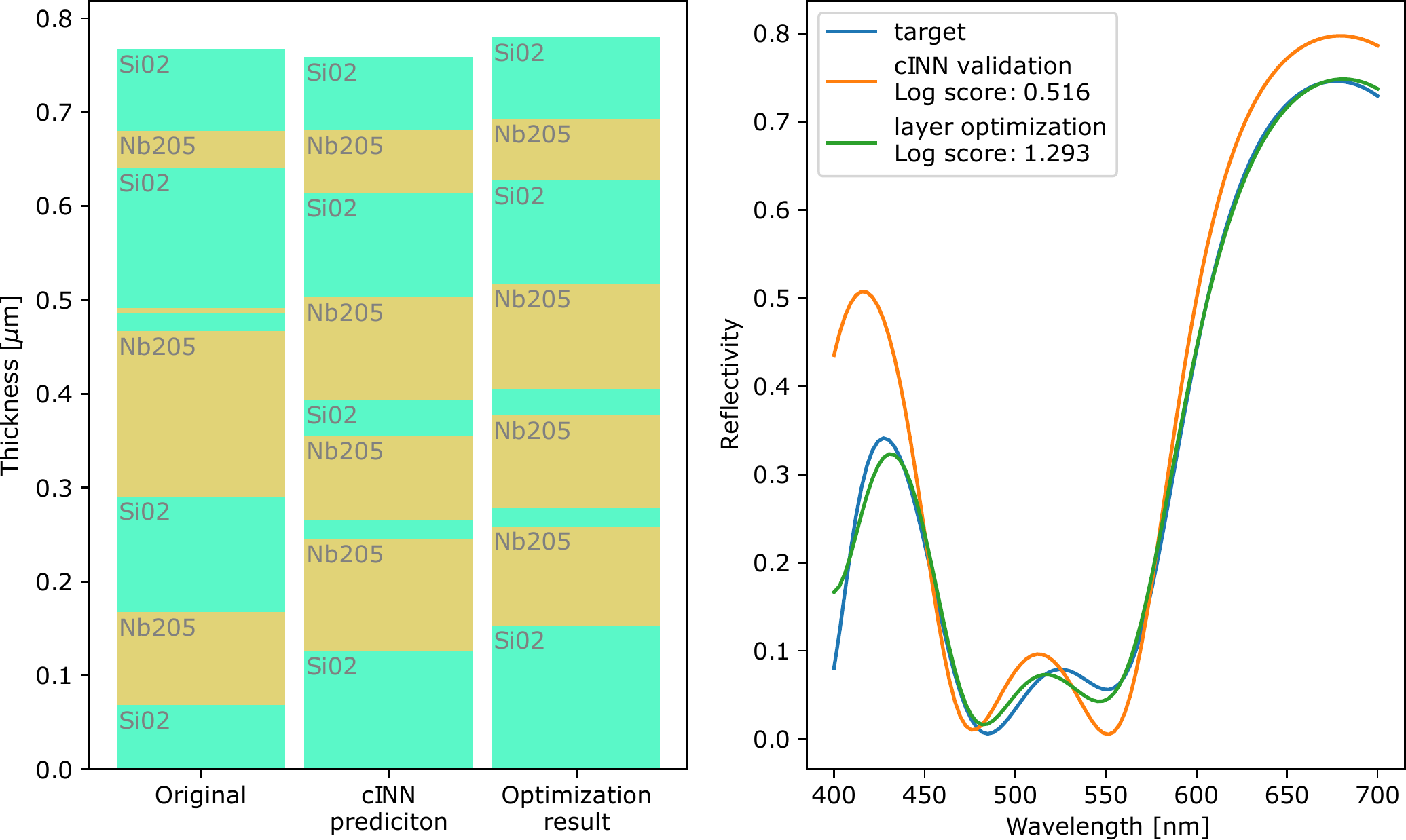}
\caption{A prediction from the cINN (center thin-film) is optimized by the Nelder-Mead downhill-simplex algorithm until convergence. By performing a subsequent optimization on the thin-films predicted by the cINN, the log-score to the target is improved significantly. The leftmost thin-film depicts the thin-film in the test dataset which was used to compute the target reflectivity. Note that the end result of the optimization deviates significantly from the original thin-film while having a similar reflectivity, underlining the present ambiguity of the optimization problem. \label{fig:test_dataset_optimization}}

\end{figure}

Applying the sketched out algorithm on the target shown in \autoref{fig:test_generative_capabilities} results in selecting \textit{Stack 1} from the four proposed thin-films for further refinement. The optimization yields thin-films with a much better approximation of the target with $\mathcal{L}_\t{ls} =1.293$. The results of one such optimization is shown in \autoref{fig:test_dataset_optimization}. Repeating the algorithm with different targets yields similar results.

\begin{figure}[t]
\centering
\includegraphics[angle=0, trim = 0.cm .0cm 0.cm 0.cm, clip, width=0.85\textwidth]{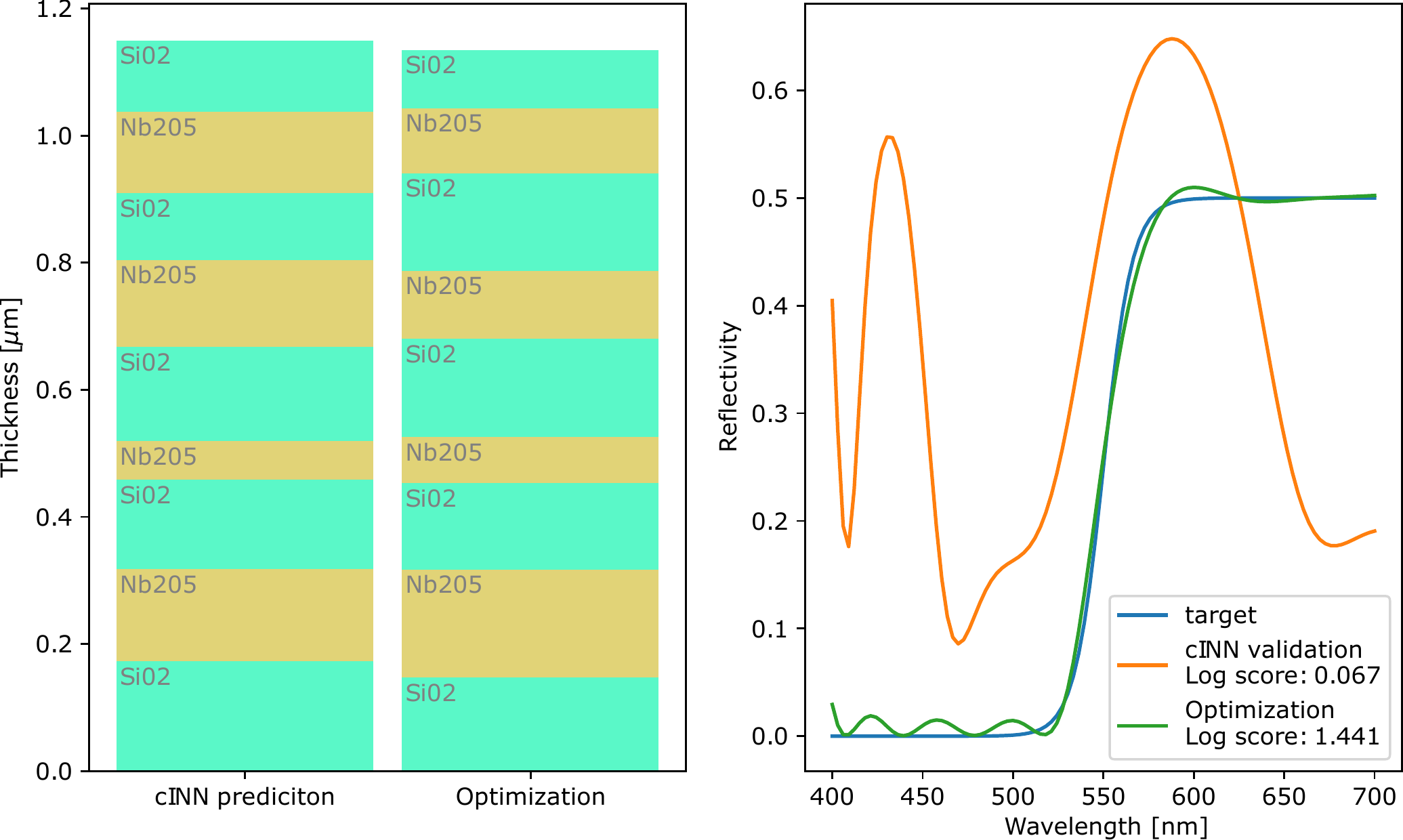}
\caption{For this optimization, a sample from the target dataset is drawn. Note that the target reflectivities might be far out-of-distribution of the training data and no known ground-truth thin-film is known a priori. By using the cINN prediction as a starting point for the downhill-simplex optimization, a thin-film that closely matches the target reflectivity is reached with $\mathcal{L}_\t{ls} =1.441$. \label{fig:target_dataset_optimization}}
\vspace{-.4cm}
\end{figure}

We can explore if the cINN can also propose thin-films for previously unseen targets by selecting a reflectivity from the target dataset. Algorithm \ref{alg:cinn_optimization} is employed and the results are shown in \autoref{fig:target_dataset_optimization}. An interesting observation is that even though the log-score of the reflectivity of the predicted thin-film is very poor ($\mathcal{L}_\t{ls} =0.067$), the thin-films of the prediction and after the optimization are very similar.

\subsection{Statistical investigation of the optimization initialization via cINN}\label{sec:statistical_investigation}
\begin{figure}[ht]
\centering
\includegraphics[angle=0, trim = 0cm .5cm 0cm 1.8cm, clip, width=1.0\textwidth]{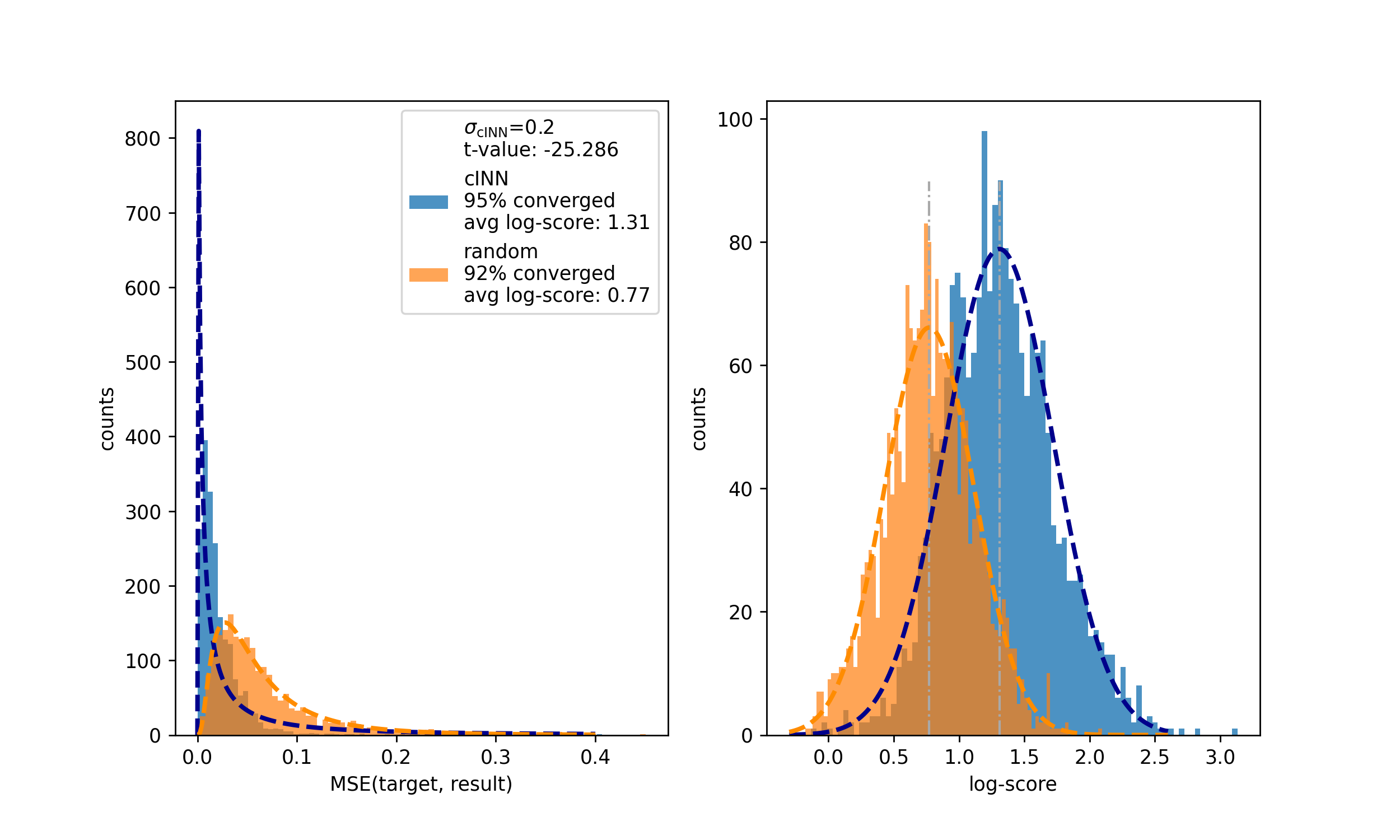}
\vspace{-.2cm}
\caption{Results of the statistical analysis. On the left is a histogram of the distribution of the mean squared error of the thin-film reflectivity of the optimization result with the optimization target reflectivity. On the right is the same distribution but measured via the log-score. The results measured by the log-score follow a Gaussian distribution approximately. These distributions are fitted with a Gaussian distribution (colored dashed line). The gray dashed line indicates the average log-score. On average, the initialization via cINN results in thin-films with better log-scores and more of the optimizations have converged. A student's t-test t-value of $t=-25.286$ indicates a statistically significant improvement over random initialization of the optimization. \label{fig:cINN_vs_random_test_dataset_simplex}}
\vspace{-.2cm}
\end{figure}

Since the cINN can't achieve the desired accuracy on its own and requires a subsequent optimization, it is important to investigate if and how much of an advantage is gained by using a cINN as initialization of the optimization. To test this, 2000 reflectivities are sampled from the test dataset. These reflectivities are used as targets for two optimizations, once with initialization via the cINN and once by sampling thin-films from the same distribution with which the training dataset was generated, see \autoref{sec:datasets}. We chose to only evaluate a sampled reflectivity once in order to prevent the selection bias of the cINN. The results of the optimization with a sampling standard deviation of $\sigma =0.2$ for the latent space is shown in \autoref{fig:cINN_vs_random_test_dataset_simplex}. On average, by initialization via the cINN the optimization converges more often, with less iterations and to better local minima than an initialization with random initial values. The same experiment was repeated with standard deviations $\sigma_i = [0.05, 0.2, 0.5, 1.0, 1.5, 2.0]$. For all of the evaluated standard deviations, the cINN initialization was superior than random initialization. The results with $\sigma=0.2$ showed the biggest improvement over random initialization while $\sigma=2.0$ showed the least improvement with an average log-score of $\mathcal{L}_\t{ls}=1.00$. Additionally, Student's t-test is performed which yields a value of $-25.286$, which suggests that the results for the presented analysis are significant beyond a reasonable doubt.

Next, the same statistical investigation was performed on all 600 target dataset reflectivities. Here, initialization of the optimization via cINN resulted in better thin-films as well. However, since many of the target reflectivities lie out-of-distribution of the training dataset and might not even be achievable given the particular material choice, the average log-score is slightly closer to the random initialization compared to the investigation of the test dataset reflectivities shown in \autoref{fig:cINN_vs_random_test_dataset_simplex}. Again, setting the standard deviation to $\sigma=0.2$ yields the best results while any initialization with the cINN is superior than using random initialization. The results are shown in \autoref{fig:cINN_vs_random_target_dataset_simplex} in the appendix.

\begin{figure}[H]
\centering
\includegraphics[angle=0, trim = 0cm 0cm 0cm 1.6cm, clip, width=0.95\textwidth]{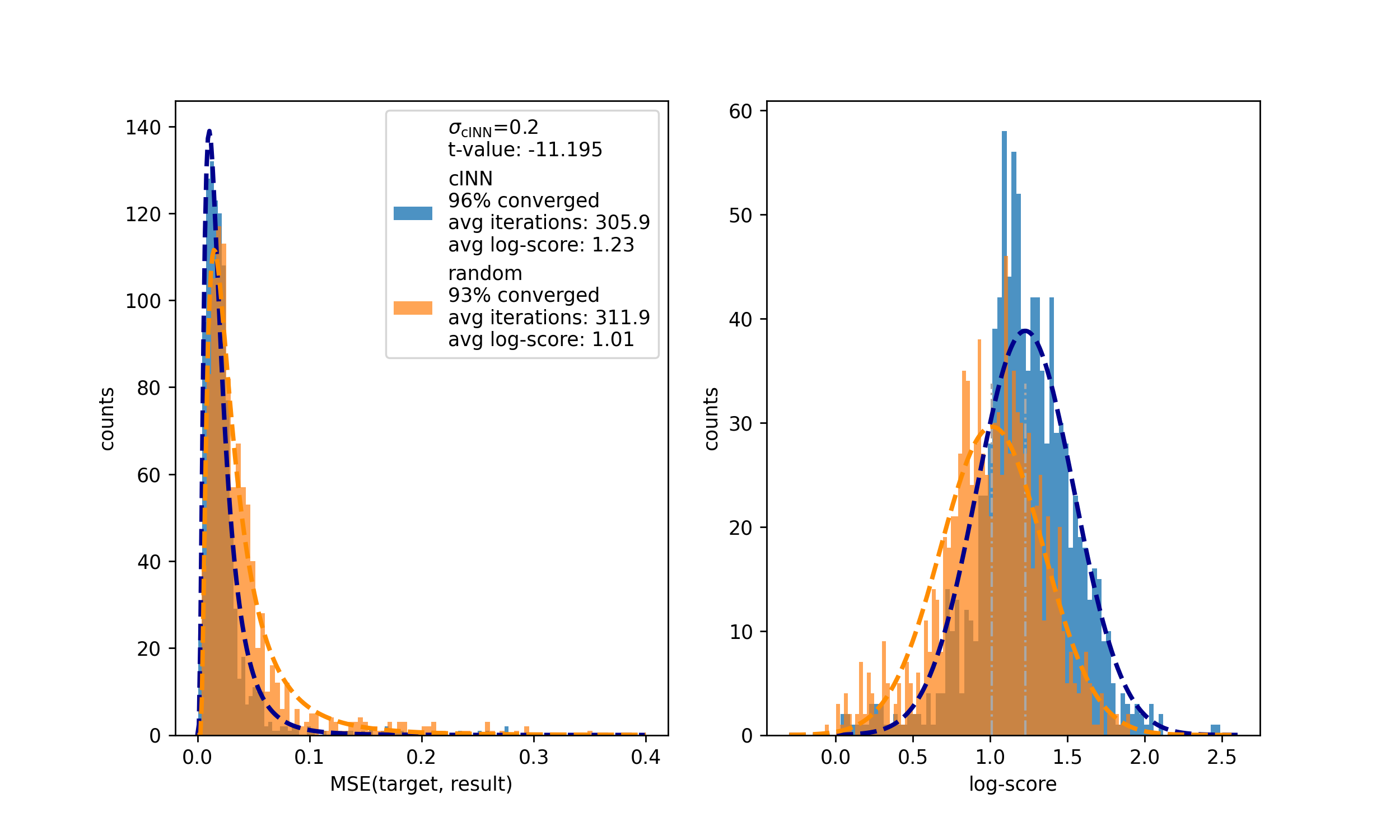}
\caption{Comparison of the results of the optimization of the same target with 1000 different random starting thin-films and 1000 starting thin-films from the cINN. Again, sampling starting points from the cINN yields better results from the optimization on average than random.  \label{fig:1000_opt_simplex}}
\vspace{-.2cm}
\end{figure}

Finally, a single reflectivity from the test dataset was chosen randomly and 1000 initial thin-films are sampled via the cINN and from the same distribution with which the training dataset was generated. Then, the initial thin-films were optimized and the found local minima are evaluated via log-score and mean squared error. The results are shown in \autoref{fig:1000_opt_simplex}. Note, that one would expect the optimizer to converge to the same local minimum for a low sampling standard deviation of the cINN. Due to the gradient approximation of the downhill-simplex algorithm, the optimization might not always converge to exactly the same local minimum if the initialisations differ slightly. 

\subsection{Comparison to state-of-the-art and discussion}
\begin{figure}[ht]
\centering
\includegraphics[angle=0, trim = 1cm .5cm 1cm 1.4cm, clip, width=1.0\textwidth]{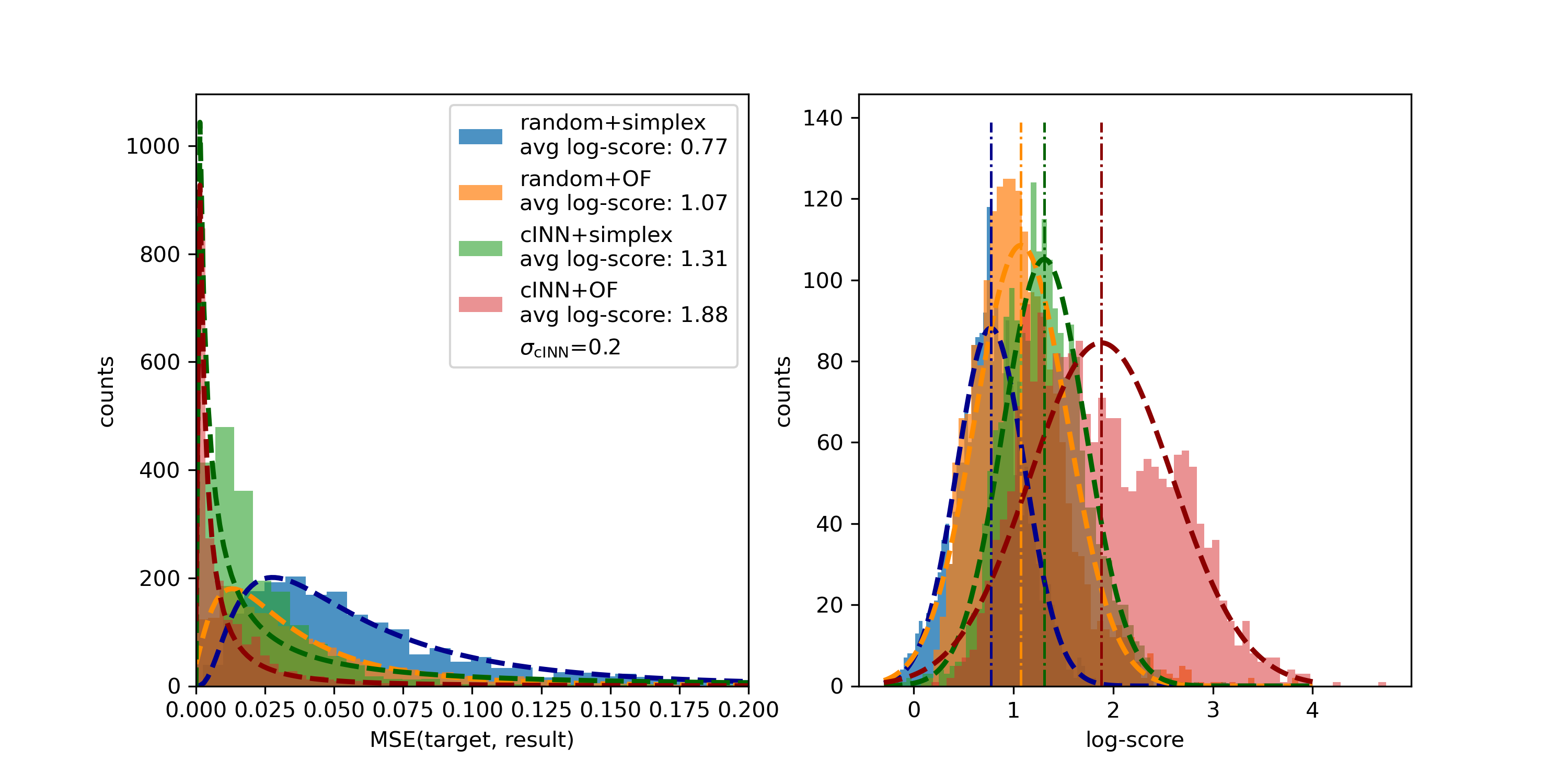}
\caption{Comparison of optimizations with cINN and random initialization by downhill simplex and OpenFilters as local optimizer. Initializing with the cINN is on average always beneficial. The minima found via the cINN initialization and optimization via downhill simplex returns better results on average than OpenFilters when used with random starting points. Using the cINN and OpenFilters achieves the best results on average. \label{fig:all_results}}
\vspace{-.2cm}
\end{figure}

There are many existing software solutions which offer an automated design process to find thin-films according to some specifications. For a comparison with state-of-the-art, we chose the Software OpenFilters (OF) \cite{Larouche_open_filters}, which designs thin-films via a two-step process. First, an initial thin-film must be given by the user. Then a refinement based on the Levenberg-Marquardt (LM) algorithm is applied to optimize the thin-film layers. If the LM algorithm has converged, the needle-point method is used and a position to introduce a new layer is chosen \cite{Sullivan_needle_point}, from which the LM algorithm continues to refine the thin-film. Here, the needle-point method is used to circumvent the problem of local minima at the expense of introducing additional layers and increasing the overall thickness of a thin-film. We chose to turn off the needle-point method in the following analysis since thin-films with more than nine layers would be created. Then we repeated the statistical analysis presented in \autoref{sec:statistical_investigation}. The results are virtually the same as with the downhill-simplex algorithm, but with a higher baseline due to the fine-tuned refinement algorithm of OpenFilters. Again, using the cINN with a sampling standard deviation of $\sigma=0.2$ yielded the best results overall while any initialization via the cINN resulted in better optimization than random initialization on average. All results for the statistical analysis of 2000 test dataset reflectivities with standard deviation $\sigma=0.2$ are shown in \autoref{fig:all_results}. Crucially, the initialization via the cINN improves the optimization via OpenFilters as well. 

The analysis show that initialization via a cINN is always advantageous compared to randomly starting a local optimization. The cINN learns to identify promising regions of thin-film parameterizations from which a local optimization converges to a local minimum. The local minima reached from the cINN propositions are very often better than starting an optimization with random parameters. This advantage can be alleviated with expert knowledge. If an expert has extensive understanding of the problem and is able to identify a promising initialization the advantage of employing a cINN may be diminished. However, it might still be interesting to query a cINN to discover promising new parameterizations. Importantly, the cINN is able to extrapolate to out-of-distribution problems as demonstrated by the target dataset reflectivities. 
Generally, the advantage of employing a cINN is greatest if the target of the problem changes frequently. Since the cINN learns the entire probability distribution of the input parameters, one obtains the biggest advantage if the entire loss landscape is also of interest. If the problem consists of finding the best possible parameters for a single target, there might by other means to obtain local minima closer to the global minimum. Wankerl et. al. showed this, by employing Reinforcement learning on a similar problem \cite{wankerl2020parameterized}. This approach lacks the option to change targets quickly though. 
Finally, we realized that the best results can be achieved if Machine Learning is used in tandem with conventional optimization techniques. This leads to much better local minima concerning the task of finding suitable thin-films with respect to some target reflectivity. It is therefore instructive to view Machine Learning as extension of generative capabilities and to exploit the best of both worlds.

\section{Conclusion}
\label{sec:5_conclusion}
In this work, we presented the application of conditional invertible neural networks on the task of generating multilayer thin-films by providing a target reflectivity. By sampling points in a Gaussian latent space, and providing a target as a condition, a cINN yields a thin-film prediction according to the probability density of the sampled latent space points. Since the learned probability distribution of the thin-film configuration is only an approximation of the true distribution, a cINN doesn't suffice to give satisfactory estimations out-of-the-box. A subsequent optimization via the downhill-simplex algorithm or by OpenFilters improved the proposed thin-films notably. By comparing the optimization results of randomly selected thin-film initialization with initialization by the cINN, we showed a significant improvement of the optimization accuracy on average by employing a cINN. 

More generally, the promising results obtained for this specific application indicate that cINN can be a very effective tool for speeding up inverse design and optimization problems, in optics and beyond. Depending on the type of problem researchers are challenged with, training a cINN to propose starting points for a subsequent optimization might be worth the initial overhead of generating a dataset and neural network training, especially if the task involves changing the optimization target frequently.









\bibliographystyle{unsrt}  
\bibliography{0_references}  

\section{Appendix}
\label{sec:appendix}

\subsection{Additional results}
\begin{figure}[H]
\vspace{-.0cm}
\centering
\includegraphics[angle=0, trim = 0cm 0cm 0cm 1.6cm, clip, width=0.95\textwidth]{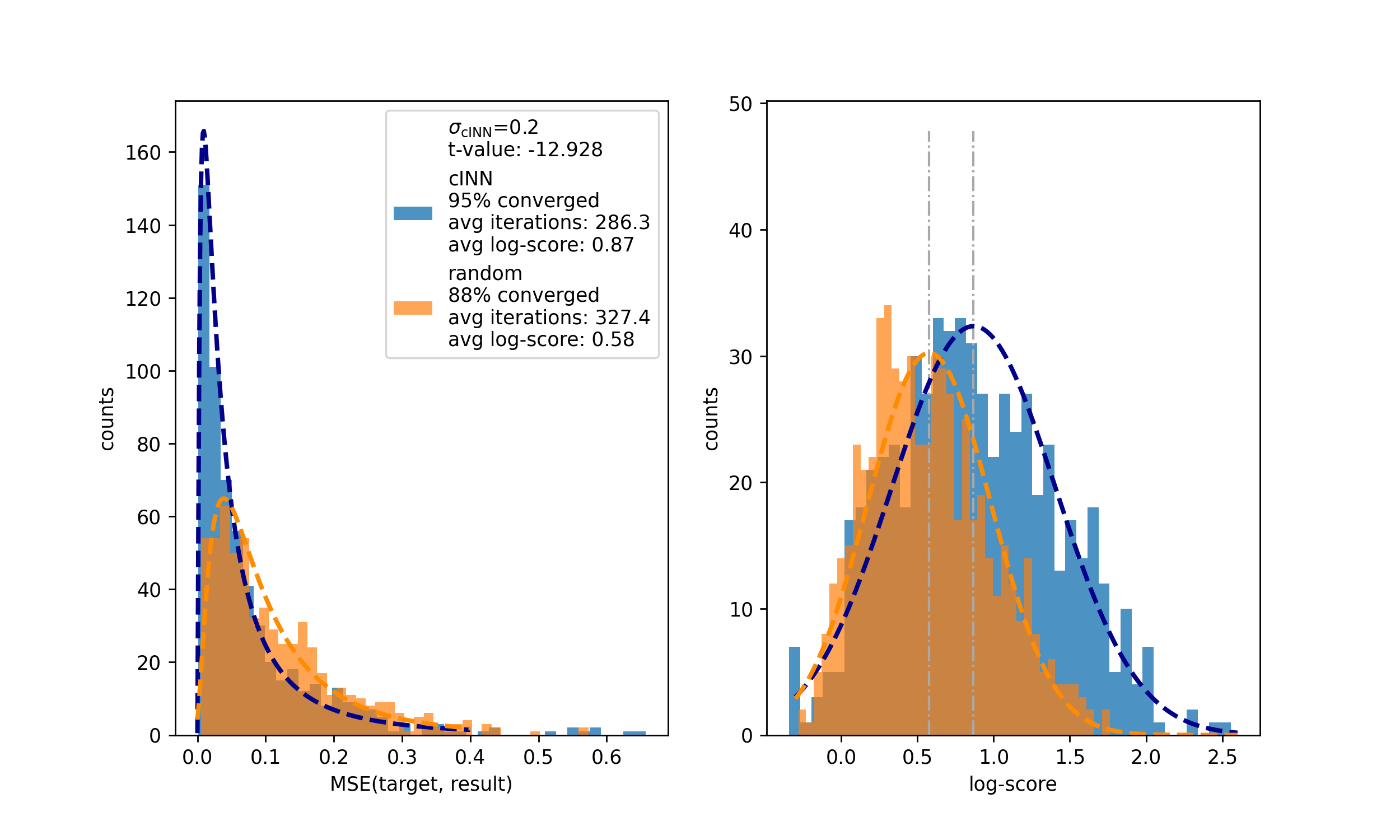}
\caption{Results of the investigation of the reflectivities of the target dataset. Here, initialization with the cINN shows significantly better log-scores on average compared to random initialization. Some of the targets are far outside of the scope of the training data and some are probably unachievable with the selected materials. Therefore, the log-scores of this investigations are worse than the log-scores of the investigation with the validation dataset reflectivities shown in \autoref{fig:cINN_vs_random_test_dataset_simplex}.  \label{fig:cINN_vs_random_target_dataset_simplex}}
\vspace{-.2cm}
\end{figure}

\end{document}